\documentstyle[12pt,aasms4]{article}
\def\be{\begin{equation}}
\def\ee{\end{equation}}
\def\ba{\begin{eqnarray}}
\def\ea{\end{eqnarray}}

\def\la{\mathrel{\mathpalette\fun <}}

\def\fun#1#2{\lower3.6pt\vbox{\baselineskip0pt\lineskip.9pt
        \ialign{$\mathsurround=0pt#1\hfill##\hfil$\crcr#2\crcr\sim\crcr}}}

\slugcomment{To be submitted to ApJ}

\begin{document}
\null\vspace{-62pt}
\begin{flushright}
astro-ph/9907405\\
October 22, 1999\\
ApJ, in press (2000)
\end{flushright}
\title{Flux-averaging Analysis of Type Ia Supernova Data}

\author{Yun Wang\footnote{Present address: Dept. of Physics, 
225 Nieuwland Science Hall, University of Notre Dame, Notre Dame, 
IN 46556-5670. email: Yun.Wang.92@nd.edu}}
\affil{{\it Princeton University Observatory} \\
{\it Peyton Hall, Princeton, NJ 08544\\}
{\it email: ywang@astro.princeton.edu}}

\vspace{.4in}
\centerline{\bf Abstract}
\begin{quotation}

Because of flux conservation, flux-averaging {\it justifies} the use
of the distance-redshift relation for a smooth universe
in the analysis of type Ia supernova (SN Ia) data.
We have combined the SN Ia data from the
High-$z$ SN Search and the Supernova Cosmology Project, and binned
the combined data by flux-averaging in redshift intervals
of $\Delta z=0.05$ and $\Delta z=0.1$.
We find that the unbinned data yield a Hubble constant
of $H_0=65\pm 1\,$km$\,$s$^{-1}\,$Mpc$^{-1}$ (statistical error only), 
a matter density fraction of $\Omega_m=0.7\pm0.4$, and a vacuum energy 
density fraction of $\Omega_{\Lambda}=1.2\pm0.5$.
The binned data for $\Delta z=0.1$ yield 
$H_0=65\pm 1\,$km$\,$s$^{-1}\,$Mpc$^{-1}$ (statistical error only), $\Omega_m=0.3\pm0.6$, 
and $\Omega_{\Lambda}=0.7\pm0.7$. 
Our results are not sensitive to the redshift bin size.
Flux-averaging leads to less biased estimates of the cosmological parameters
by reducing the bias due to systematic effects such as weak lensing.

Comparison of the data of 18 SNe Ia published by both groups yields 
a mean SN Ia peak absolute magnitude of $M_B=-19.33\pm 0.25$.
The internal dispersion of each data set is about 0.20 magnitude
in the {\it calibrated} SN Ia peak absolute magnitudes.
The difference in analysis techniques introduces an additional
uncertainty of about 0.15 magnitude. 

If the SNe Ia peak absolute luminosity changes with redshift
due to evolution, our ability to measure the cosmological parameters
from SN Ia data will be significantly diminished.
Assuming power-law evolution in the peak absolute luminosity, $(1+z)^{\beta}$,
we find a strong degeneracy between the evolution power-law 
index $\beta$ and the matter density fraction $\Omega_m$.
For $\Omega_m=0.3$, we find that the unbinned data yields
$H_0=65\pm 1\,$km$\,$s$^{-1}\,$Mpc$^{-1}$ (statistical error only), 
$\Omega_{\Lambda}=1.4\pm 1.1$, and $\beta=0.5\pm 1.6$,
and the binned data (with $\Delta z=0.1$) yields
$H_0=65\pm 1\,$km$\,$s$^{-1}\,$Mpc$^{-1}$ (statistical error only), 
$\Omega_{\Lambda}=0.6\pm 1.4$, and $\beta=0.0\pm 1.0$.

\end{quotation}


\section{Introduction}

Near the end of the millennium, cosmology has become a data-driven 
science. We are closer than ever to determining the fundamental cosmological
parameters which describe our observable universe 
(\cite{neta99,Eisen99,Mike99,WSS99}).
The use of astrophysical standard candles provides a fundamental means
of measuring the cosmological parameters $H_0$ (current expansion rate
of the universe), $\Omega_m$ (matter density fraction of the universe), 
and $\Omega_{\Lambda}$ (vacuum energy density fraction of the universe).

Type Ia supernovae (SNe Ia) are currently our best candidates for 
standard candles. They can be calibrated to have small intrinsic 
dispersions (\cite{Phillips93,Riess95}).
Two independent teams, the High-$z$ SN Search (Schmidt et al.) and
the Supernova Cosmology Project (Perlmutter et al.), have observed about
100 SNe Ia. The data analysis results of both teams seem to 
suggest that our universe has a significant vacuum energy content
(\cite{Garna98,Riess98,Perl99}).

In this paper, we combine the data of the High-$z$ SN Search team and
the Supernova Cosmology Project, and bin the combined data by
flux-averaging in redshift intervals. Previous work (\cite{Wang99b})
has shown that flux-averaging is effective in reducing the scatter
in SN Ia peak absolute luminosity due to weak gravitational lensing
and intrinsic dispersions. Here, we study the effect of flux-averaging
on the estimation of cosmological parameters.
In \S 2 we compare the data sets from the two teams.
In \S 3 we describe how we combine the data from the two teams.
In \S 4 we flux-average the combined data and derive estimated
cosmological parameters.
In \S 5 we illustrate the effect of evolution of SN peak absolute 
luminosity on the estimation of cosmological parameters.
\S 6 contains a summary.

\section{Comparison of data sets}

The published data of the High-$z$ SN Search team 
(\cite{Schmidt98,Riess98}) consists of 50 SNe Ia.
They give measured distance modulus for each SN Ia, $\mu_0$, to be compared 
with the theoretical prediction
\be
\label{eq:mu0p}
\mu_0^p= 5\,\log\left( \frac{ d_L}{\mbox{Mpc}} \right)+25,
\ee
where $d_L(z)$ is the luminosity distance; it is
related to the angular diameter distance $d_A(z)$
and comoving distance $r(z)$:
\be
\label{eq:dL}
d_L(z)=(1+z)^2 d_A(z)=(1+z)\, r(z),
\ee
assuming a completely smooth universe.

The comoving distance $r(z)$ is given by (\cite{Weinberg72})
\be
\label{eq:r(z)}
r(z)=\frac{cH_0^{-1}}{|\Omega_k|^{1/2}}\,
\mbox{sinn}\left\{ |\Omega_k|^{1/2}
\int_0^z dz'\,\left[ \Omega_m(1+z')^3+\Omega_{\Lambda}+\Omega_k(1+z')^2
\right]^{-1/2} \right\},
\ee
where ``sinn'' is defined as sinh if $\Omega_k>0$, and sin if  $\Omega_k<0$.
If $\Omega_k=0$, the sinn and $\Omega_k$'s disappear from Eq.(\ref{eq:r(z)}),
leaving only the integral.

The published data of the Supernova Cosmology Project 
(\cite{Perl99}) consists of 60 SNe Ia.
They give the estimated effective B-band magnitude of each SN Ia, 
to be compared with
\be
\label{eq:convert}
m_B^{eff}= M_B+\mu_0^p,
\ee
where $M_B$ is the peak absolute magnitude of a ``standard'' SN Ia
in the B-band, and $\mu_0^p$ is given by Eq.(\ref{eq:mu0p}).

We find that the published data of the two teams have 
18 SNe Ia in common. The two data sets for the same SNe Ia should differ
by the constant offset $M_B$. We find
\ba
\label{eq:dispMLCSPerl}
M_B^{MLCS} &\equiv & m_B^{eff}-\mu_0^{MLCS}=-19.33\pm 0.25, \\
M_B^{m15} &\equiv & m_B^{eff}-\mu_0^{m15}=-19.42\pm 0.27, 
\ea
where $\mu_0^{MLCS}$ and $\mu_0^{m15}$ for each SN Ia are
estimated using the MLCS method and the template-fitting method (m15)
respectively by Riess et al. (1998).
The difference in the data from the two teams has a dispersion
of about 0.25 magnitudes. This is surprisingly large since 16 of these
SNe Ia were drawn from the Hamuy et al. (1996) data.
For these 16 SNe Ia, we find
$ m_B^{eff}-\mu_0^{MLCS}=-19.30\pm 0.24$, and
$m_B^{eff}-\mu_0^{m15}=-19.40\pm 0.26$.
For the rest of this section, we only consider these 16 SNe Ia.

The simplest way to calibrate SNe Ia is to use the linear relation
between maximum peak luminosity and decline time found by Phillips (1993).
Hamuy et al. (1996) found $M^B_{max}=a+b[\Delta m_{15}(B)-1.1]$,
with $a=-19.258(0.048)$ and $b=0.784(0.182)$
for 26 ``low extinction'' SNe Ia.
Table 1 lists 16 SNe Ia from Hamuy et al. (1996) data
that have been reanalyzed by both Riess et al. (1998)
and Perlmutter et al. (1999).
\begin{table}[htb]
\caption{16 SNe Ia from Hamuy et al. (1996)}
\begin{center}
\begin{tabular}{c|c|c|c|c|c|c|c}
\hline\hline
SN Ia	& $z$ & $M_B^{H96}$ & $m_B^{eff}-\mu_0^{MLCS}$ &
$m_B^{eff}-\mu_0^{m15}$ & $\mu_0^{MLCS}-\mu_0^p$ & $\mu_0^{m15}-\mu_0^p$ &
$m_B^{eff}-\mu_0^p$
\\
\hline
1990af &   0.050 &   -19.31  & -18.90  & -19.04 &  -0.39  &  -0.25 &   -19.29\\
1992P  &   0.026 &   -19.16  & -19.68  & -19.51 &   0.31  &   0.14 &  -19.37\\
1992ae &   0.075 &   -19.21  & -19.37  & -19.34 &  -0.05  &  -0.08 &  -19.42\\
1992ag &   0.026 &   -19.05  &  -19.09 &  -19.25&    -0.08&   0.08 &  -19.17\\
1992al &   0.014 &   -19.48  & -19.45 &  -19.66&    -0.16&   0.05 &  -19.61\\
1992aq &   0.101 &   -19.17  &  -19.25 &  -19.17&    -0.14&  -0.22 &  -19.39\\
1992bc &   0.020 &   -19.46  &  -19.69 &  -19.59&     0.00&  -0.10 &  -19.69\\
1992bg &   0.035 &   -19.40  &  -19.60 &  -19.83&     0.14&   0.37 &  -19.46\\
1992bh &   0.045 &   -18.85  &  -19.30 &  -19.26&     0.23&   0.19 &  -19.07\\
1992bl &   0.043 &   -19.45  &  -19.07 &  -19.34&    -0.32&  -0.05 &  -19.39\\
1992bo &   0.018 &   -19.22  &  -19.11 &  -19.27&     0.08&   0.24 &  -19.03\\
1992bp &   0.080 &   -19.57  & -19.38 &  -19.69&    -0.35&  -0.04 &  -19.73\\
1992br &   0.087 &   -19.12  &  -18.93 &  -18.81&     0.01&  -0.11 &  -18.92\\
1992bs &  0.064  &  -18.98   & -19.37  &  -19.39&     0.12&   0.14 &  -19.25\\
1993O  &   0.052 &   -19.32  &  -19.49 &  -19.77&     0.02&   0.30 &  -19.47\\
1993ag &   0.050 &   -19.27  &  -19.11 &  -19.42&    -0.12&   0.19 &  -19.23\\
\hline
\hline
\end{tabular}
\end{center}
\end{table}

In Table 1,
$M_B^{H96}$ is the {\it corrected} B band peak absolute magnitude  
given by Hamuy et al. (1996):
\be
\label{eq:MBH96}
M_B^{H96} \equiv M_{max}^B+5\log(H_0/65)
-0.784\,(\Delta m_{15}-1.1).
\ee
We find that for the 16 SNe Ia listed, $M_B^{H96}=-19.253\pm 0.190$.

From Table 1, we find
\ba
\label{eq:comH96}
(m_B^{eff}-\mu_0^{MLCS})-M_B^{H96} &= &
-0.047 \pm 0.270 \nonumber\\
(m_B^{eff}-\mu_0^{m15})-M_B^{H96} &= &
-0.144 \pm 0.229.
\ea
This indicates that the absolute magnitude
of a ``standard'' SN Ia derived from comparing the Riess et al. (1998)
and the Perlmutter et al. (1999) data sets differs appreciably from that
derived by Hamuy et al. (1996) from the same 16 SNe Ia.

To account for these differences, we examine the internal dispersion
of each data set. Using $\mu_0^p$ from Eq.(\ref{eq:mu0p}) with 
$z\ll 1$ and $H_0=65\,$km$\,$s$^{-1}\,$Mpc$^{-1}$,
we find
\ba
\label{eq:dispint}
\mu_0^{MLCS}-\mu_0^p &=&  -0.043 \pm 0.195 \nonumber\\
\mu_0^{m15}-\mu_0^p &=&  0.054 \pm 0.178 \nonumber\\
m_B^{eff}-\mu_0^p &=&  -19.34 \pm 0.22
\ea
These should be compared with $M_B^{H96}=-19.253\pm 0.190$ for the 16 
SNe Ia calibrated using Eq.(\ref{eq:MBH96}).
Fig.1 shows the internal dispersions in the calibrated SN Ia 
peak absolute magnitudes,
as given by (a) $\mu_0^{MLCS}-\mu_0^p$, (b) $\mu_0^{m15}-\mu_0^p$,
(c) $m_B^{eff}-\mu_0^p$, and (d) $M_B^{H96}$.
Clearly, the internal dispersion for each data set 
is about $\sigma_{int}\sim 0.20$ magnitude.
This indicates that these 16 SNe Ia can be calibrated to be standard
candles with a dispersion (intrinsic and observational) of about 0.2 magnitude.

Next, we examine how the difference in analysis techniques introduces
uncertainty. For the Riess et. al 1998
data analyzed using MLCS method and template-fitting method 
respectively, we find
\be
\mu_0^{MLCS}-\mu_0^{m15}=-0.097 \pm 0.163.
\ee
For the Perlmutter et al. (1999) data and the Hamuy et al. (1996) data,
we find
\be
(m_B^{eff}-\mu_0^p)-M_B^{H96}=-0.09 \pm 0.147.
\ee
It seems that the difference in analysis technique typically introduces
a uncertainty of $\sigma_{tech}\sim 0.15$ magnitude.

Therefore, the difference between independently analyzed data sets could be 
as large as $\sigma = \sqrt{\sigma_{int}^2+\sigma_{tech}^2} \sim 0.25$.
The differences between the Riess et al. (1998) data and
the Perlmutter et al. (1999) data stem mostly from the internal
dispersion in the peak absolute magnitudes of SNe Ia
estimated from each data set, with a substantial
contribution from the difference in analysis techniques.

Eq.(\ref{eq:comH96}) indicates that the SN Ia absolute magnitude estimated
using $m_B^{eff}-\mu_0^{MLCS}$ is closer 
to that found by Hamuy et al. (1996).
We will use the MLCS method data from Riess et al. (1998) for the
rest of this paper.

\section{Combination of Data}

We combine the data from the two teams by adding 42 SNe Ia from
the Supernova Cosmology Project to the High-$z$ SN Search data
(the MLCS method results),
leaving out the 18 SNe Ia from the Supernova Cosmology Project 
which are already included in the High-$z$ SN Search data.
This yields a combined data set of 92 SNe Ia. 
In the combined data set, we convert $m_B^{eff}$ of the 
the SNe Ia which have been taken from the Supernova Cosmology Project 
to $\mu_0^{MLCS}$ by using Eq.(\ref{eq:convert}) with $M_B=-19.33$.
$M_B=-19.33$ is the mean difference in the data $(m_B^{eff}-\mu_0^{MLCS})$
of the 18 SNe Ia published
by both teams [see Eq.(\ref{eq:dispMLCSPerl})]. Fig.2 shows the
difference between $m_B^{eff}$ and $\mu_0^{MLCS}$ for these 18 SNe Ia.

Fig.3 shows the magnitude-redshift plots of the combined data set of 
92 SNe Ia. The solid points represent 50 SNe Ia from Schmidt et al. data.
The circles represent 42 additional SNe Ia from Perlmutter et al. 
The dashed line is the prediction of the open cold dark matter model 
(OCDM) with $\Omega_m=0.2$ and $\Omega_{\Lambda}=0$. 
The dotted line is the prediction
of the standard cold dark matter model (SCDM) with $\Omega_m=1$
and $\Omega_{\Lambda}=0$. The two thick solid lines are flat models
with a cosmological constant ($\Lambda$CDM):
($\Omega_m, \,\Omega_{\Lambda})=(0.3,\,0.7)$ and
($\Omega_m, \,\Omega_{\Lambda})=(0.2,\,0.8)$.
The thin solid line is the SCDM model 
($\Omega_m, \,\Omega_{\Lambda})=(1,\,0)$
with $(1+z)$ dimming of SN Ia peak luminosity (linear evolution).
Fig.3(a) shows the distance modulus versus redshift.
Fig.3(b) shows the distance modulus relative to the prediction
of the OCDM model.

\subsection{Estimation of the cosmological parameters}

We find that the best-fit values of the cosmological parameters are
sensitive to the estimated errors of the data points.
Following Riess et al. (1998), we use
\be
\chi^2(H_0,\Omega_m, \Omega_{\Lambda})=
\sum_i \frac{ \left[ \mu^p_{0,i}(z_i| H_0,\Omega_m, \Omega_{\Lambda})-
\mu_{0,i} \right]^2 }{\sigma_{\mu_0,i}^2 +\sigma_{mz,i}^2},
\ee
where $\sigma_{\mu_0}$ is the estimated measurement error of the distance
modulus, and $\sigma_{mz}$ is the dispersion in the distance modulus 
due to the dispersion in galaxy redshift, $\sigma_z$, due to
peculiar velocities and uncertainty in the galaxy redshift
(for the Perlmutter et al. data, the dispersion due to peculiar
velocities is included in $\sigma_{m_B^{eff}}$, i.e., $\sigma_{\mu_0}$).
Since
\be
\label{eq:sigmamz}
\sigma_{mz}=\frac{5}{\ln 10} \left( 
\frac{1}{d_L}\frac{\partial d_L}{\partial z} \right)\, \sigma_z,
\ee
$\sigma_{mz}$ depends on $\Omega_{\Lambda}$ and $\Omega_m$. We compute
$\sigma_{mz}$ iteratively while estimating the cosmological parameters.

For the Schmidt et al. data, we follow Riess et al. (1998) in 
adopting $\sigma_z=200\,$km$\,$s$^{-1}$, and add 2500 km$\,$s$^{-1}$
in quadrature to $\sigma_z$ for SNe Ia whose redshifts were determined 
from broad features in the SN spectrum. For the Perlmutter et al. data,
we take $\sigma_{\mu_0}=\sigma_{m_B^{eff}}$, which already includes
dispersion due to peculiar velocities of 300$\,$km$\,$s$^{-1}$, and
we use the redshift uncertainty $\sigma_z$ for each SN Ia given in their 
tables to compute $\sigma_{mz}$ (see Eq.(\ref{eq:sigmamz})).

Table 2 lists the derived best-fit cosmological parameters, 
with 1-$\sigma$ error 
bars, for the High-$z$ SN Search data (Schmidt et al.),  
the Supernova Cosmology Project data (Perlmutter et al.), and 
the combined data. Note that $h$ is the dimensionless Hubble constant, 
defined by $H_0 =100\, h\,$km$\,$s$^{-1}$Mpc$^{-1}$.
$\chi^2_{\nu}$ is $\chi^2$ per degree of freedom.
\begin{table}[htb]
\caption{Estimated cosmological parameters with SNe Ia}
\begin{center}
\begin{tabular}{c|c|c|c}
\hline\hline
	& 50 SNe Ia (Schmidt et al.) & 60 SNe Ia (Perlmutter et al.) &
92 SNe Ia (combined data)\\
\hline
$h$ & 0.65 $\pm$ 0.01$^*$ & 0.66 $\pm$ 0.02$^*$ & 0.65 $\pm$ 0.01$^*$\\
$\Omega_m$ & 0.2 $\pm$ 0.6 & 1.0 $\pm$ 0.4 & 0.7 $\pm$ 0.4\\
$\Omega_{\Lambda}$ & 0.7 $\pm$ 0.8 & 1.7 $\pm$ 0.6 & 1.2 $\pm$ 0.5\\
$\chi^2_{\nu}$ & 1.16 & 1.63 & 1.48 \\
\hline
\end{tabular}
\tablecomments{*Statistical error only, not including the contribution
from the much larger SN Ia absolute magnitude error.}
\end{center}
\end{table}

Fig.4 shows the 68.3\% and 95.4\% confidence contours in the 
$\Omega_{\Lambda}-\Omega_m$ plane. The dotted lines represent
the Schmidt et al. data (50 SNe Ia), the dashed lines represent
the Perlmutter et al. data (60 SNe Ia), and the solid lines represent
the combined data (92 SNe Ia).
Note that we have allowed $\Omega_m$ to be a free parameter in
estimating its value. Since distances are {\it not} directly measurable,
we can treat them as theoretical intermediaries in the data analysis,
therefore $\Omega_m$ must be allowed to have negative as well
as positive and zero values, for statistical robustness.

We have combined data from two independent teams analyzed using 
different statistical methods.
It is important to analyze all the SN data with the same techniques
for statistical consistency; this is not possible at present, because
not all the reduced SN Ia data have become public
(Perlmutter et al. have not yet published the light curves of their 42 SNe Ia).
However, the published data from the two teams are similar enough
(with a constant offset) to allow a meaningful study of the effect
of flux-averaging on the estimation of cosmological parameters 
from SN Ia data (see \S 4 and \S 5). 

\subsection{$H_0$ as a systematic indicator of data sets}

We have chosen to estimate $H_0$ simultaneously with $\Omega_m$ and $\Omega_{\Lambda}$
mainly for three reasons. 

First, $H_0$ is a useful indicator of the systematic ``zero point'' of a given SN Ia
data set. Let us consider the 16 SNe Ia from Hamuy et al. (1996) that both teams
(Riess et al. 1998, Perlmutter et al. 1999) have reanalyzed (see \S 2).
Since each of these 16 SNe Ia can only have one true peak absolute magnitude,
we take $M_B^{MLCS}=M_B^{eff}=M_B^{H96}$, where $M_B^{MLCS}$, $M_B^{eff}$,
and $M_B^{H96}$ are the peak absolute magnitudes derived from the 
Riess et al. 1998 (MLCS), Perlmutter et al. 1999, and Hamuy et al. 1996 data respectively.
From Eq.(\ref{eq:dispint}),  
$\mu_0^{MLCS}-\mu_0^p = -0.043$ and $m_B^{eff}-\mu_0^p =  -19.34$.
Using $m_B^{eff}=M_B^{eff}+\mu_0^{eff}$, and setting $M_B^{eff}=M_B^{H96}=-19.253$,
we find $\mu_0^{eff}-\mu_0^p =-0.087$. Hence
\be
\mu_0^{MLCS}-\mu_0^{eff}=-\frac{5}{\ln 10} \frac{\Delta H_0}{H_0}=0.044,
\ee
i.e., the difference between the estimated values for the
Hubble constant from Riess et al. 1998
($H_0^{MLCS}$) and Perlmutter et al. 1999 ($H_0^{eff}$) data is 
\be
H_0^{MLCS}-H_0^{eff}=-1.32,
\ee
for $H_0^{MLCS}=65$.
This is in agreement with the estimated values of $H_0$ 
listed in Table 2 for 50 SNe Ia from the
Riess et al. (1998) data and for 60 SNe Ia from the Perlmutter et al. (1999) data.
The SNe Ia from Perlmutter et al. (1999) are systematically brighter than
the Riess et al. (1998) SNe Ia by about 0.04 magnitude.

Second, integrating out the $H_0$ dependence in the probability distribution 
functions of the cosmological parameters have very little effect on the 
estimated values of $\Omega_m$ and $\Omega_{\Lambda}$, because the 
$H_0$ dependence is uncorrelated with the $\Omega_m$ and $\Omega_{\Lambda}$
dependence. Therefore, we have little to gain in the accuracy of estimated 
$\Omega_m$ and $\Omega_{\Lambda}$ by excluding $H_0$ from the parameter 
estimation.

Thrid, including $H_0$ in the parameter estimation conforms with the 
common practice of estimating all basic cosmological parameters
simultaneously when analyzing other cosmological data sets, for example, the Cosmic
Microwave Background Anisotropy data (\cite{WSS99,Eisen99}).
The distance modulus depends on $H_0$, as well as
$\Omega_m$ and $\Omega_{\Lambda}$. Therefore, $H_0$ should be estimated from the SNe Ia
data simultaneously with $\Omega_m$ and $\Omega_{\Lambda}$, although the $H_0$ dependence
of the data is independent of the dependence on $\Omega_m$ and $\Omega_{\Lambda}$.

To summarize, including $H_0$ in the cosmological parameter estimation 
from SN Ia data provides a useful systematic indicator of the data set 
used, while having little effect on the accuracy of the other cosmological 
parameters ($\Omega_m$ and $\Omega_{\Lambda}$) being estimated, and it 
conforms with the convention used in parameter estimation from other 
cosmological data.
Note that the error of the $H_0$ estimated here only reflects statistical
error, it does not include the contribution from the much larger SN Ia 
peak absolute magnitude error. For a realistic assessment of the actual 
value and errors associated with $H_0$ estimated from SNe Ia, the reader 
is referred to Saha et al. (1999), Jha et al. (1999), and 
Gibson et al. (1999).

\section{Flux-averaging of data}

An important reason to consider flux-averaging is the weak gravitational
lensing of SNe Ia. Because our universe is inhomogeneous in matter 
distribution, weak gravitational lensing leads to a non-Gaussian 
distribution in the magnification of standard candles.

For a given redshift $z$, if a mass-fraction $\tilde{\alpha}$ of the matter 
in the universe is smoothly distributed, the largest possible distance for
light bundles which have not passed through a caustic is given by
the solution to the Dyer-Roeder equation (\cite{DR73,Sch92,Kantow98}):
\ba
\label{eq:DR}
&& g(z) \, \frac{d\,}{dz}\left[g(z) \frac{dD_A}{dz}\right]
+\frac{3}{2} \tilde{\alpha} \,\Omega_m (1+z)^5 D_A=0, \nonumber \\
&& D_A(z=0)=0, \hskip 1cm \left.\frac{dD_A}{dz}\right|_{z=0}=\frac{c}{H_0},
\ea
where $g(z) \equiv (1+z)^3 \sqrt{ 1+ \Omega_m z+ \Omega_{\Lambda}
[(1+z)^{-2} -1] }$. The smoothness parameter $\tilde{\alpha}$
essentially represents the amount of matter that can cause the
magnification of a given source.
If we define a direction-dependent smoothness parameter $\tilde{\alpha}$ 
via the Dyer-Roeder equation, there is a unique mapping between
$\tilde{\alpha}$ and the magnification of a source.
We can think of our universe as a mosaic of cones centered on
the observer, each with a different value of $\tilde{\alpha}$.
Wang (1999) has derived empirical fitting formulae for the
probability distribution of $\tilde{\alpha}$ from the numerical
simulation results of Wambsganss et al. (1997).

Wang (2000) showed that flux-averaging of simulated data (with noise
due to both weak lensing and intrinsic dispersions) leads to 
SN peak luminosities which well approximate the true luminosities
with $\tilde{\alpha}=1$ (which represents a completely smooth universe).
The angular-diameter distance defined
in Eq.(\ref{eq:dL}) using the comoving distance $r(z)$, 
$d_A(z)=r(z)/(1+z)$, satisfies Eq.(\ref{eq:DR}) with
$\tilde{\alpha}=1$, i.e., the angular-diameter distance defined
in Eq.(\ref{eq:dL}) is the Dyer-Roeder distance with
$\tilde{\alpha}=1$. Therefore, flux-averaging {\it justifies} the use
of Eq.(\ref{eq:dL}) in the analysis of SN Ia data.

Before flux-averaging, we convert the distance modulus of SNe Ia into 
``fluxes'', $f(z_i)=10^{-\mu_0(z_i)/2.5}$. We then obtain ``absolute 
luminosities'', \{${\cal L}(z_i)$\}, by
removing the redshift dependence of the ``fluxes'', i.e.,
\be
{\cal L}(z_i) = 4\pi\,d_L^2(z_i|H_0,\Omega_m, 
\Omega_{\Lambda})\,f(z_i),
\ee
where $d_L$ is the luminosity distance, and $(H_0,\Omega_m, \Omega_{\Lambda})$ 
are the best-fit cosmological parameters derived from the unbinned data
set \{$f(z_i)$\}. We then flux-average over the 
``absolute luminosities'' \{${\cal L}_i$\} in each redshift bin.
The set of best-fit cosmological parameters derived from the binned data 
is applied to the unbinned data \{$f(z_i)$\} to obtain a new set 
of ``absolute luminosities'' \{${\cal L}_i$\}, which is then flux-averaged 
in each redshift bin, and the new binned data is used to derive a 
new set of best-fit cosmological parameters. This procedure is repeated until
convergence is achieved.
The 1-$\sigma$ error on each binned data point is taken to be the
root mean square of the 1-$\sigma$ errors on the unbinned data points
in the given redshift bin, \{$f_i$\} ($i=1,2,...,N$), 
multiplied by $1/\sqrt{N}$ (see \cite{Wang99b}).

Fig.5 shows magnitude-redshift plots of the binned data for 
the total of 92 SNe Ia,
with redshift bin $\Delta z=0.05$. The lines are the same as in Fig.1.
Fig.5(a) shows the distance modulus versus redshift.
Fig.5(b) shows the distance modulus relative to the prediction
of the open cold dark matter model (OCDM) with $\Omega_m=0.2$
and $\Omega_{\Lambda}=0$. 
Fig.6 is the same as Fig.5, but with redshift bin $\Delta z=0.1$.

Table 3 lists the estimated cosmological parameters, with 1-$\sigma$ error 
bars. 
\begin{table}[htb]
\caption{Estimated cosmological parameters with 92 SNe Ia}
\begin{center}
\begin{tabular}{c|ccc}
\hline\hline
	& unbinned data &  binned with $\Delta z=0.05$ &
 binned with $\Delta z=0.1$ \\
$N_{data}$ & 92 & 17 & 10 \\
\hline
$h$ & 0.65 $\pm$ 0.01$^*$ & 0.65 $\pm$ 0.01$^*$ & 0.65 $\pm$ 0.01$^*$\\
$\Omega_m$ & 0.7 $\pm$ 0.4 & 0.3 $\pm$ 0.6 & 0.3 $\pm$ 0.6\\
$\Omega_{\Lambda}$ & 1.2 $\pm$ 0.5 & 0.7 $\pm$ 0.7 & 0.7 $\pm$ 0.7\\
$\chi^2_{\nu}$ & 1.48 & 0.73 & 0.78 \\
\hline\hline
&  & fixing $\Omega_{\Lambda}=1-\Omega_m$ & \\
\hline
$h$ & 0.65 $\pm$ 0.01$^*$ & 0.65 $\pm$ 0.01$^*$ & 0.65 $\pm$ 0.01$^*$\\
$\Omega_m$ & 0.3 $\pm$ 0.1 & 0.3 $\pm$ 0.1 & 0.3 $\pm$ 0.1\\
$\chi^2$ & 1.47 & 0.68 & 0.68 \\
\hline\hline
&  & fixing $\Omega_{\Lambda}=0$ & \\
\hline
$h$ & 0.64 $\pm$ 0.01$^*$ & 0.65 $\pm$ 0.01$^*$ & 0.65 $\pm$ 0.01$^*$\\
$\Omega_m$ & -0.2 $\pm$ 0.1 & -0.2 $\pm$ 0.1 & -0.2 $\pm$ 0.1\\
$\chi^2$ & 1.49 & 0.72 & 0.76 \\
\hline\hline
&  & fixing $\Omega_m=0.3$ & \\
\hline
$h$ & 0.65 $\pm$ 0.01$^*$ & 0.65 $\pm$ 0.01$^*$ & 0.65 $\pm$ 0.01$^*$\\
$\Omega_{\Lambda}$ & 0.7 $\pm$ 0.1 & 0.7 $\pm$ 0.2 & 0.7 $\pm$ 0.2\\
$\chi^2$ & 1.47 & 0.68 & 0.68 \\
\hline
\end{tabular}
\tablecomments{*Statistical error only, not including the contribution
from the much larger SN Ia absolute magnitude error.}
\end{center}
\end{table}

Fig.7 shows the 68.3\% and 95.4\% confidence contours in the 
$\Omega_{\Lambda}-\Omega_m$ plane. The solid lines represent the unbinned
data, the dotted lines represent the binned data with 
redshift bin $\Delta z=0.05$; the dashed lines represent
the binned data with redshift bin $\Delta z=0.1$.

Fig.8 shows that the estimated parameters from flux-averaged data
are not sensitive to the size of the redshift bin. 
The thick solid lines 
are the estimated $\Omega_m$ (a) and $\Omega_{\Lambda}$ (b) as 
functions of the redshift bin size $dz$, the dotted lines mark the
1$\sigma$ errors on $\Omega_m$ (a) and $\Omega_{\Lambda}$ (b).
The thin solid line is $\chi^2_{\nu}+2$.
The wiggles in the lines are due to the small number of SNe Ia
in each redshift bin for small bin size, and due to the small number of
binned data points for large bin size.
Clearly, the optimal range for the bin size is $0.025\la dz \la 0.9$.

Because flux-averaging reduces the bias due to weak lensing,
the flux-averaged data yield less biased estimates of the 
cosmological parameters.

\section{Effect of evolution}

Drell, Loredo, \& Wasserman (1999)
have studied the effect of
SN Ia luminosity evolution on the estimation of cosmological
parameters.
Recently, Riess et al. (1999) suggested that there is an indication
of evolution of SNe Ia from their risetimes, and that if this observed
evolution affects the peak luminosities of SNe Ia, it can account
for the observed faintness of high-$z$ SNe Ia without invoking a
cosmological constant. To illustrate this interesting possibility,
we introduce a power-law evolution in the SNe Ia peak luminosity, 
${\cal L}(z)= 4\pi\,d_L^2(z)\,f(z)$, i.e.,
\be
{\cal L}(z)= (1+z)^{\beta}\, {\cal L}(z=0).
\ee
Here, $\beta>0$ represents brightening of SN Ia with redshift,
while $\beta<0$ represents dimming with redshift.
The thin solid line in Fig.3, Fig.5, and Fig.6 represent the
SCDM model ($\Omega_m=1$, $\Omega_{\Lambda}=0$) with linear dimming
of the SN Ia peak luminosity with $\beta=-1$.

Table 4 lists the estimated cosmological parameters and SN Ia evolution
power-law index $\beta$, with 1-$\sigma$ error bars, for the combined
data set of 92 SNe Ia.
\begin{table}[htb]
\caption{Estimated cosmological parameters with SN evolution}
\begin{center}
\begin{tabular}{c|ccc}
\hline\hline
	& unbinned data &  binned with $\Delta z=0.05$ &
 binned with $\Delta z=0.1$ \\
$N_{data}$ & 92 & 17 & 10 \\
\hline
$h$ & 0.66 $\pm$ 0.02$^*$ & 0.65 $\pm$ 0.02$^*$ & 0.65 $\pm$ 0.03$^*$\\
$\Omega_m$ & 1.5 $\pm$ 5.8 & 0.8 $\pm$ 22.5 & 0.9 $\pm$ 31.3\\
$\Omega_{\Lambda}$ & 1.2 $\pm$ 0.8 & 0.4 $\pm$ 4.5 & 0.4 $\pm$ 3.3\\
$\beta$ & -0.7 $\pm$ 3.7 &-0.6 $\pm$ 17.1 & -0.6 $\pm$ 21.1\\
$\chi^2_{\nu}$ & 1.49 & 0.78 & 0.91 \\
\hline\hline
&  & fixing $\Omega_m=0.3$ & \\
\hline
$h$ & 0.65 $\pm$ 0.01$^*$ & 0.65 $\pm$ 0.01$^*$ & 0.65 $\pm$ 0.01$^*$\\
$\Omega_{\Lambda}$ & 1.4 $\pm$ 1.1 & 0.7 $\pm$ 1.3 & 0.6 $\pm$ 1.4\\
$\beta$ & 0.5 $\pm$ 1.6 &0.0 $\pm$ 1.0 & 0.0 $\pm$ 1.0\\
$\chi^2$ & 1.59 & 0.73 & 0.78 \\
\hline
\end{tabular}
\tablecomments{*Statistical error only, not including the contribution
from the much larger SN Ia absolute magnitude error.}
\end{center}
\end{table}

There is strong degeneracy, as expected, between the evolution power-law 
index $\beta$ and the matter density fraction $\Omega_m$.
If the SNe Ia peak absolute luminosity changes with redshift
due to evolution, our ability to measure the cosmological parameters
from SN Ia data will be significantly diminished, unless
we can correct for the evolution.
The issue of SN evolution can only be settled if a large number of SNe Ia
are observed at $z>1$; this can be accomplished via a supernova pencil
beam survey using a dedicated 4 meter telescope (\cite{Wang99b}).

\section{Conclusions}

We have combined the data from the two independent SN Ia groups,
Schmidt et al. and Perlmutter et al., and analyzed the combined
data using flux-averaging in redshift bins of 0.05 and 0.1. 
We find that the estimation of the cosmological parameters
is not sensitive to the size of the redshift bin.
While the combined data without flux-averaging are best fit by
a closed universe with high matter content and larger than critical
density vacuum energy, the flux-averaged data are best fit by
a nearly flat universe with a low matter content. This is consistent
with the strong observational evidence that we live in a low matter density 
universe (\cite{neta95,Carlberg96,neta98,Krauss98,neta99}).
Flux-averaging leads to less biased 
estimates of the cosmological parameters by reducing the bias due to
systematic effects such as weak lensing.

The distance-redshift relation, Eq.(3), is generally used in making
theoretical predictions to compare with the brightnesses of standard
candles. However, it is only valid in a smooth universe.
We live in a clumpy universe. Because of flux conservation, flux-averaging
justifies the use of Eq.(3) in the analysis of SN Ia data (see \S 4).

In the estimation of cosmological parameters, we have allowed
the matter density fraction $\Omega_m$ and the vacuum energy density 
fraction $\Omega_{\Lambda}$ to be unconstrained free parameters.
Unlike fluxes, distances are $not$ directly measurable, therefore they
are best treated as theoretical intermediaries.
Since the cosmological parameters $\Omega_m$ and $\Omega_{\Lambda}$
enter through distances [see Eqs.(2) \& (3)], they should be allowed to have 
negative as well as positive and zero values for statistical robustness.
By not applying priors on $\Omega_m$ and $\Omega_{\Lambda}$ in our
parameter estimation, we can check the validity of Eqs.(2) \& (3).
For example, if the most probable value of $\Omega_m$ derived from
the data is negative ($\Omega_m<0$), it could be an indication that
Eqs.(2) and (3) do not apply, since we know that $\Omega_m>0$.

We have also investigated the effect of possible evolution of 
SNe Ia on the estimation of cosmological parameters. 
Assuming power-law evolution in the SN Ia peak absolute luminosity, 
we find that there is strong degeneracy between the evolution power-law 
index and the matter density in the universe.
The evolution of SNe Ia must be
resolved before they can be regarded as reliable standard candles.
Supernova pencil beam surveys which could yield hundreds of SNe Ia
at $z>1$ over a few years will be critical in constraining the evolution
of SNe Ia (\cite{Wang99b}).

\acknowledgements{\centerline{\bf Acknowledgements}}
It is a pleasure for me to thank Neta Bahcall for helpful comments
and a careful reading of the manuscript, and the referee for
useful suggestions.


\clearpage
\setcounter{figure}{0}

\figcaption[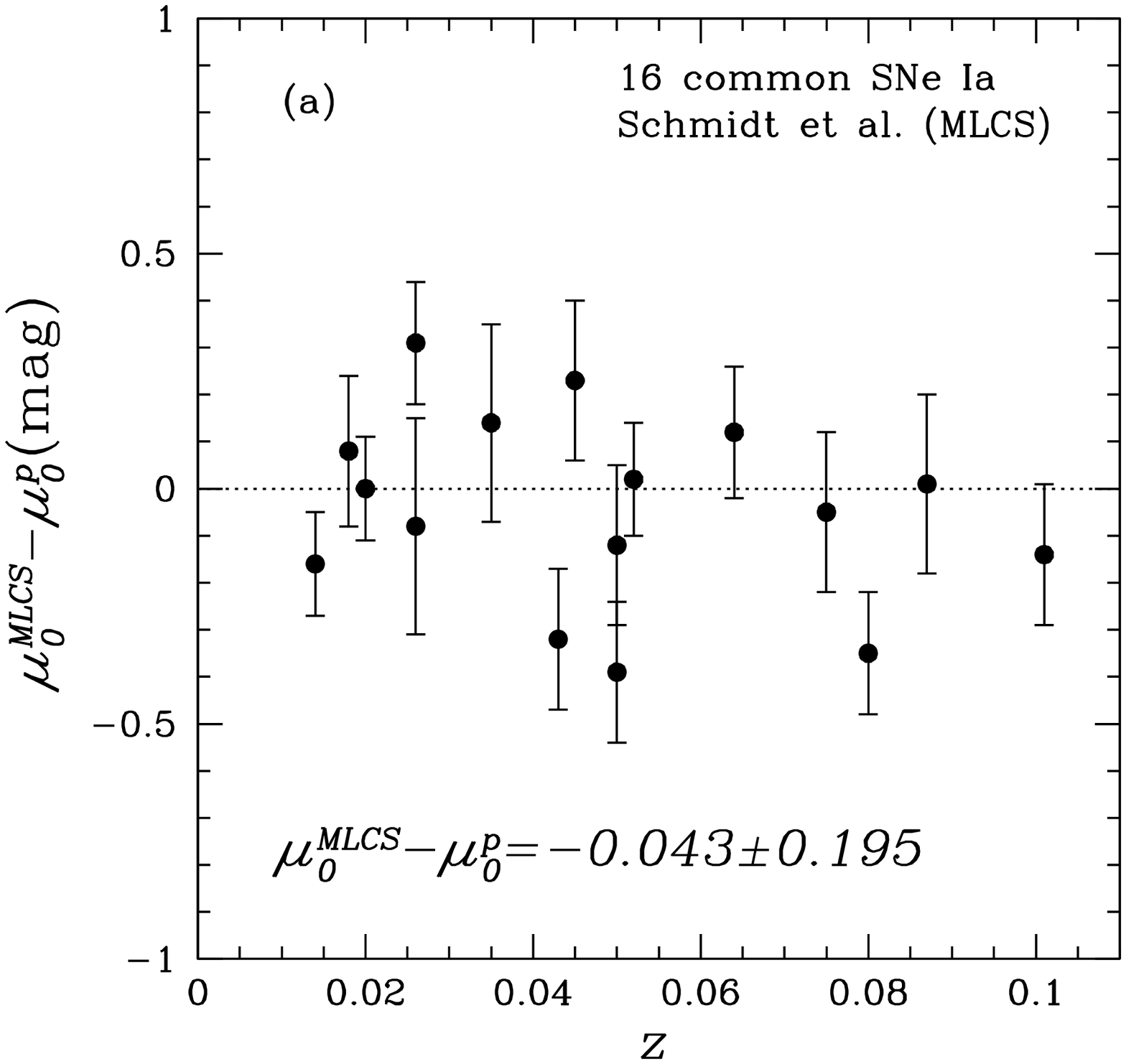]
{The internal dispersions in the calibrated SN Ia peak absolute magnitudes
from four data sets.
(a) $\mu_0^{MLCS}-\mu_0^p$, (b) $\mu_0^{m15}-\mu_0^p$,
(c) $m_B^{eff}-\mu_0^p$, and (d) $M_B^{H96}$.}

\figcaption[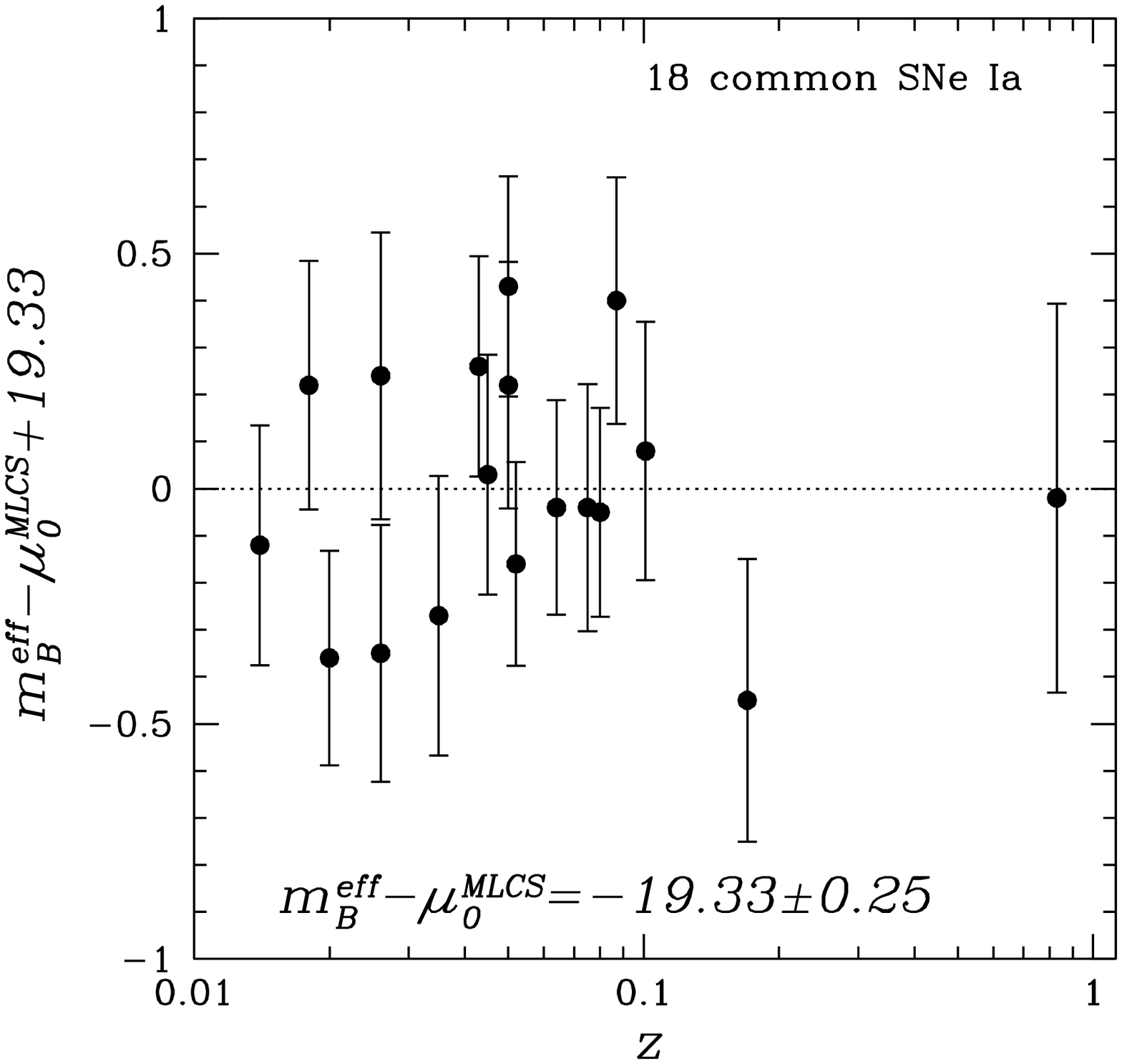]
{The difference between $m_B^{eff}$ (Perlmutter et al. 1999)
and $\mu_0^{MLCS}$ (Riess et al. 1998) for the same 18 SNe Ia.
The error bars are the combined errors in $m_B^{eff}$ and $\mu_0^{MLCS}$.} 

\figcaption[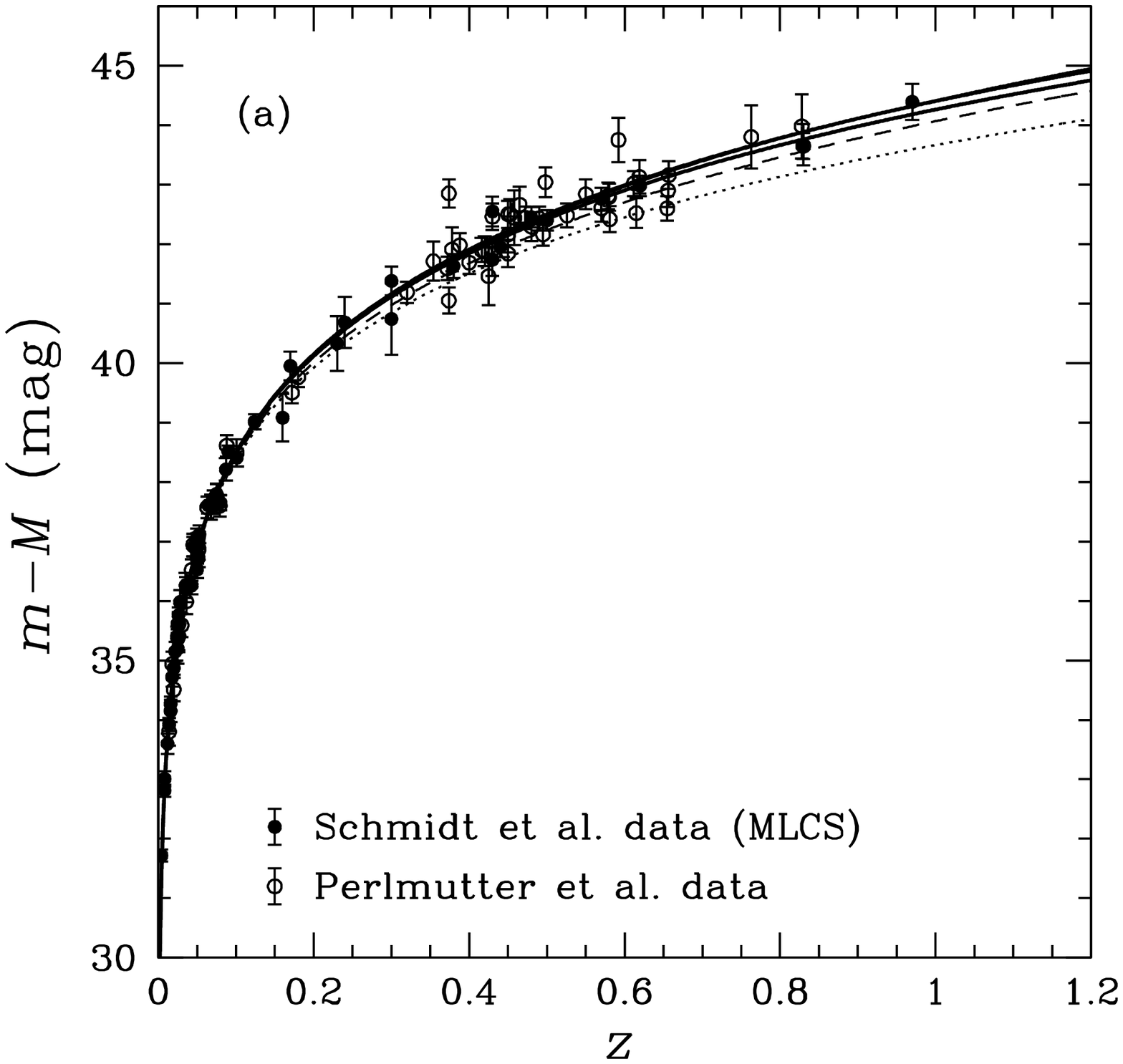]
{The magnitude-redshift plots of the combined data set of 
92 SNe Ia. The solid points represent 50 SNe Ia from Schmidt et al. data.
The empty points represent 42 additional SNe Ia from Perlmutter et al.
data. The dashed line is the prediction of OCDM with $\Omega_m=0.2$ 
and $\Omega_{\Lambda}=0$. 
The dotted line is the prediction of SCDM with $\Omega_m=1$
and $\Omega_{\Lambda}=0$. The two thick solid lines are predictions
of $\Lambda$CDM, with ($\Omega_m, \,\Omega_{\Lambda})=(0.3,\,0.7)$,
($\Omega_m, \,\Omega_{\Lambda})=(0.2,\,0.8)$ respectively.
The thin solid line is SCDM with $(1+z)$ dimming of SN Ia peak luminosity.
(a) The distance modulus versus redshift.
(b) The distance modulus relative to the prediction
of the OCDM model.}

\figcaption[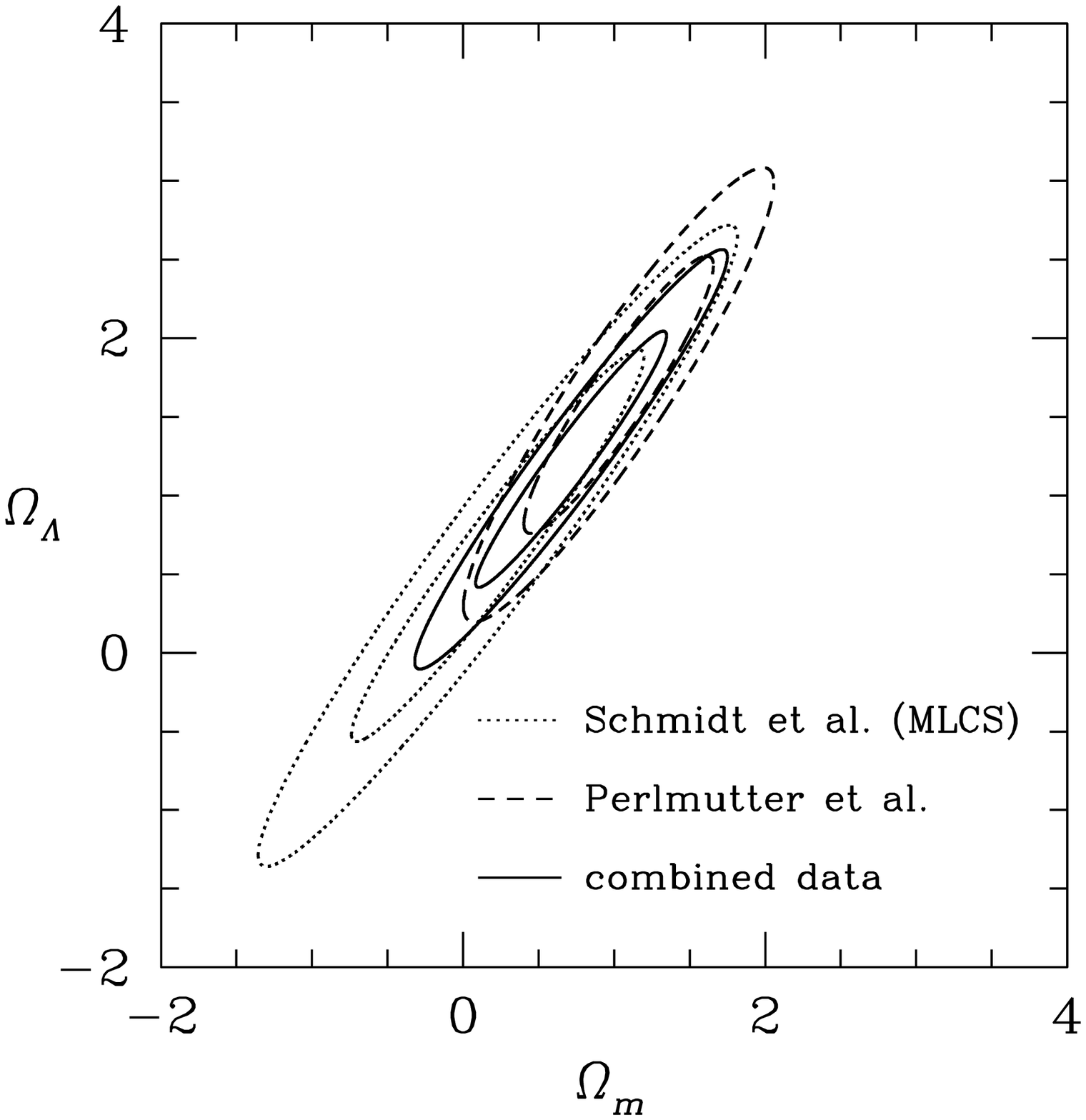]
{The 68.3\% and 95.4\% confidence contours in the 
$\Omega_{\Lambda}-\Omega_m$ plane. The dotted lines represent
the Schmidt et al. data (50 SNe Ia), the dashed lines represent
the Perlmutter et al. data (60 SNe Ia), and the solid lines represent
the combined data (92 SNe Ia).}

\figcaption[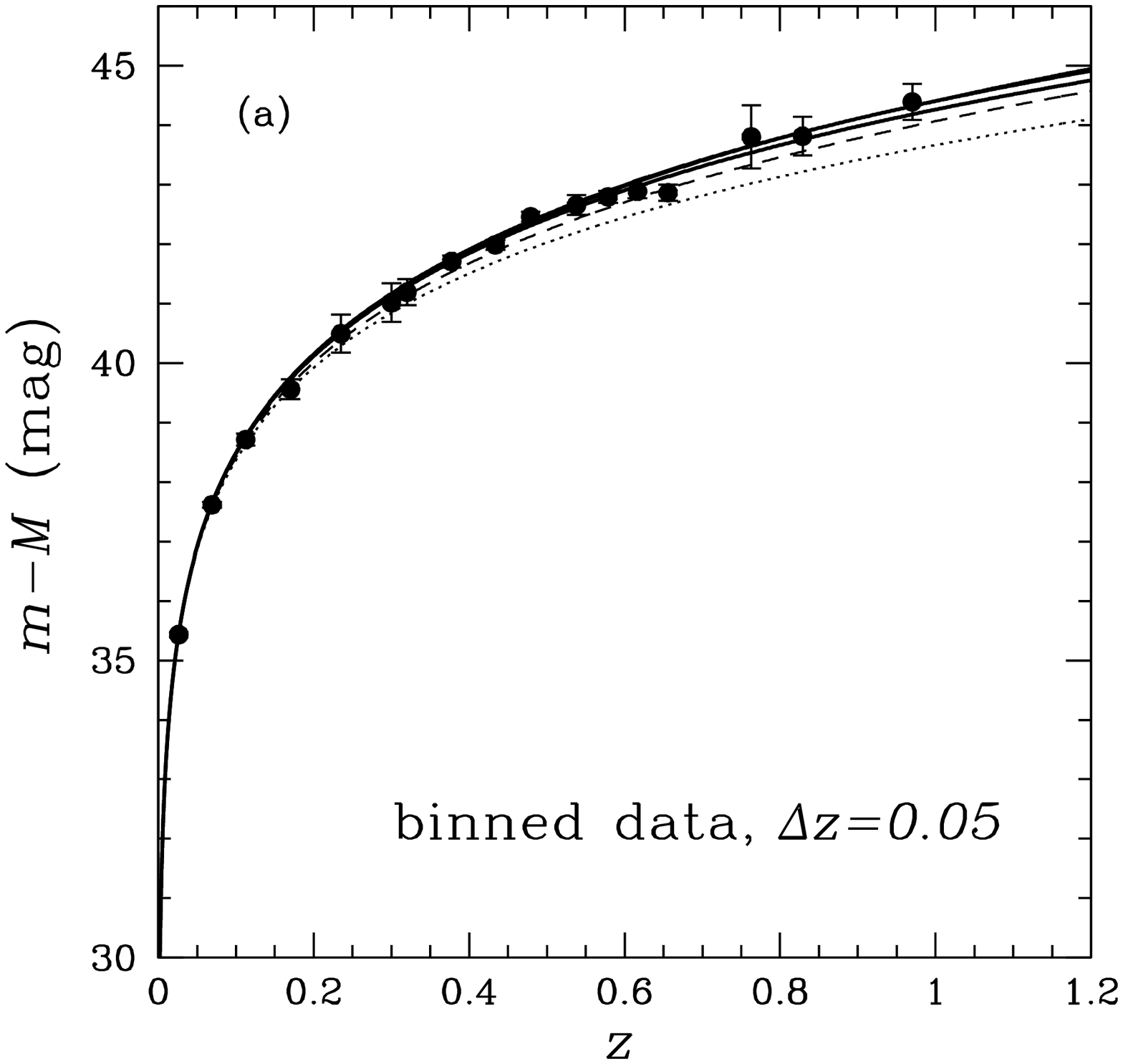]
{The magnitude-redshift plots of the binned data for 92 SNe Ia,
with redshift bin $\Delta z=0.05$.
The line types are the same as in Fig.3.
(a) The distance modulus versus redshift.
(b) The distance modulus relative to the prediction
of the open cold dark matter model (OCDM) with $\Omega_m=0.2$
and $\Omega_{\Lambda}=0$.}

\figcaption[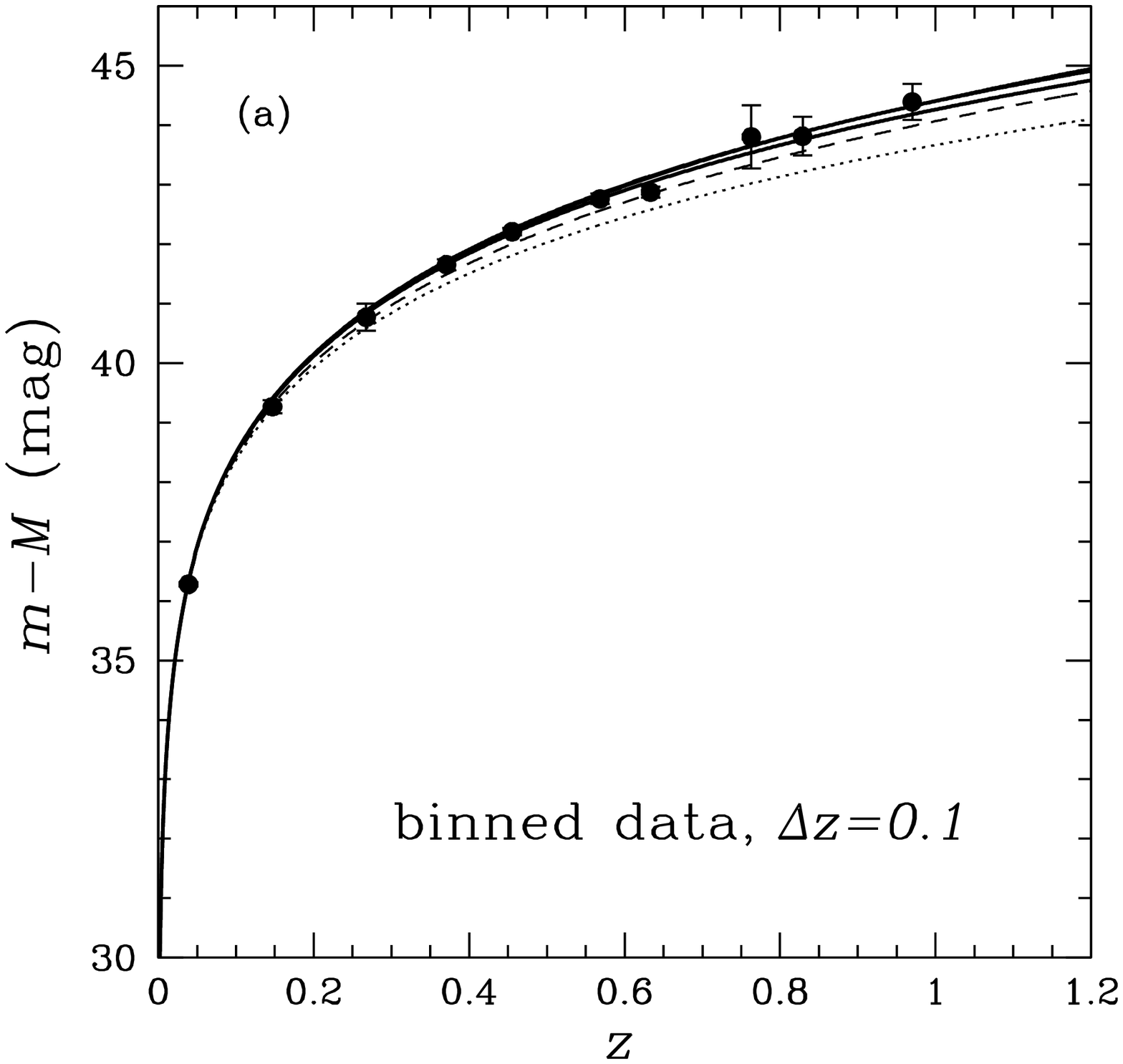]
{The same as Fig.5, but with redshift bin $\Delta z=0.1$.}

\figcaption[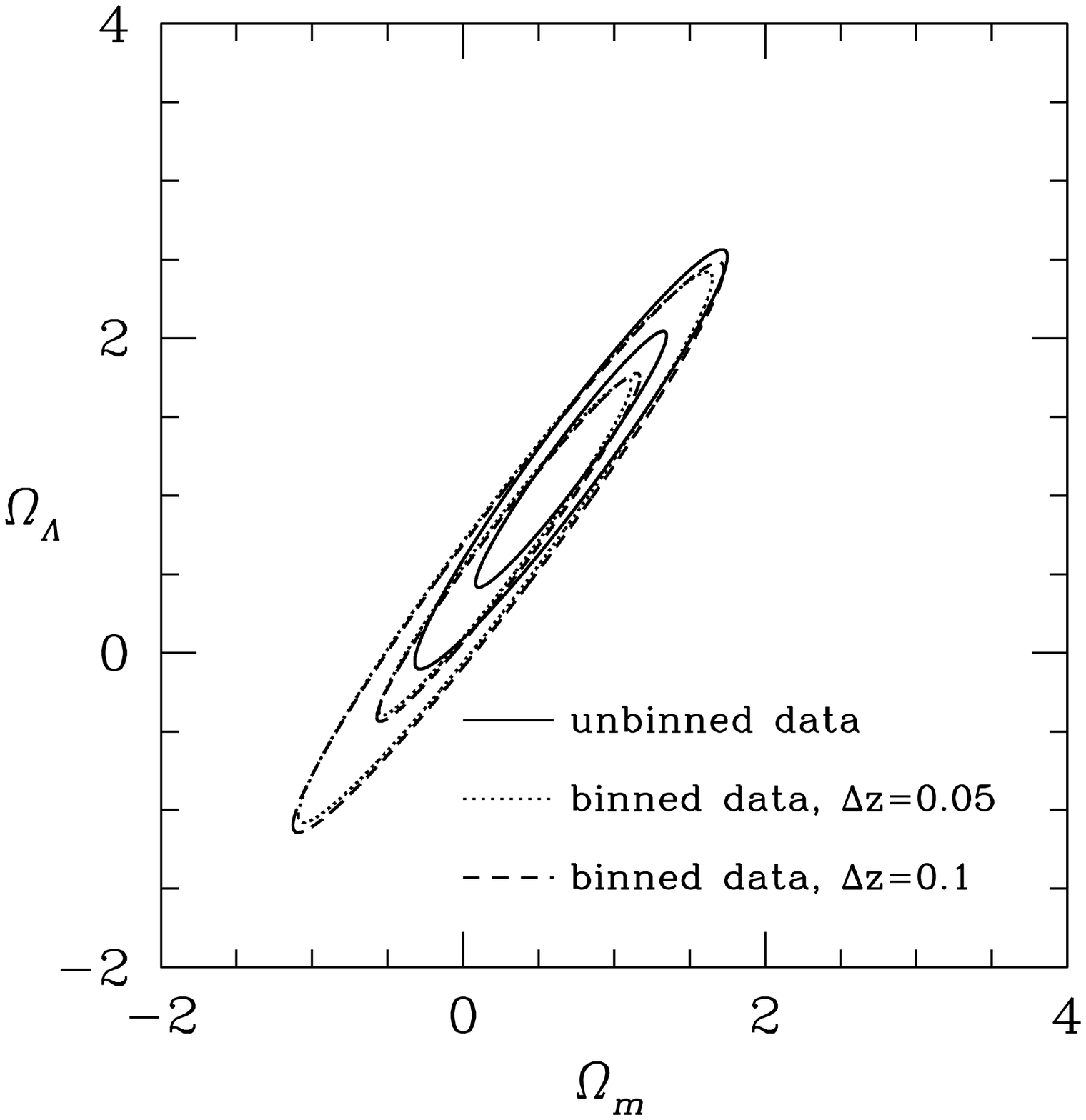]
{The 68.3\% and 95.4\% confidence contours in the 
$\Omega_{\Lambda}-\Omega_m$ plane. The solid lines represent the unbinned
data, the dotted lines represent the binned data with 
redshift bin $\Delta z=0.05$; the dashed lines represent
the binned data with redshift bin $\Delta z=0.1$.}

\figcaption[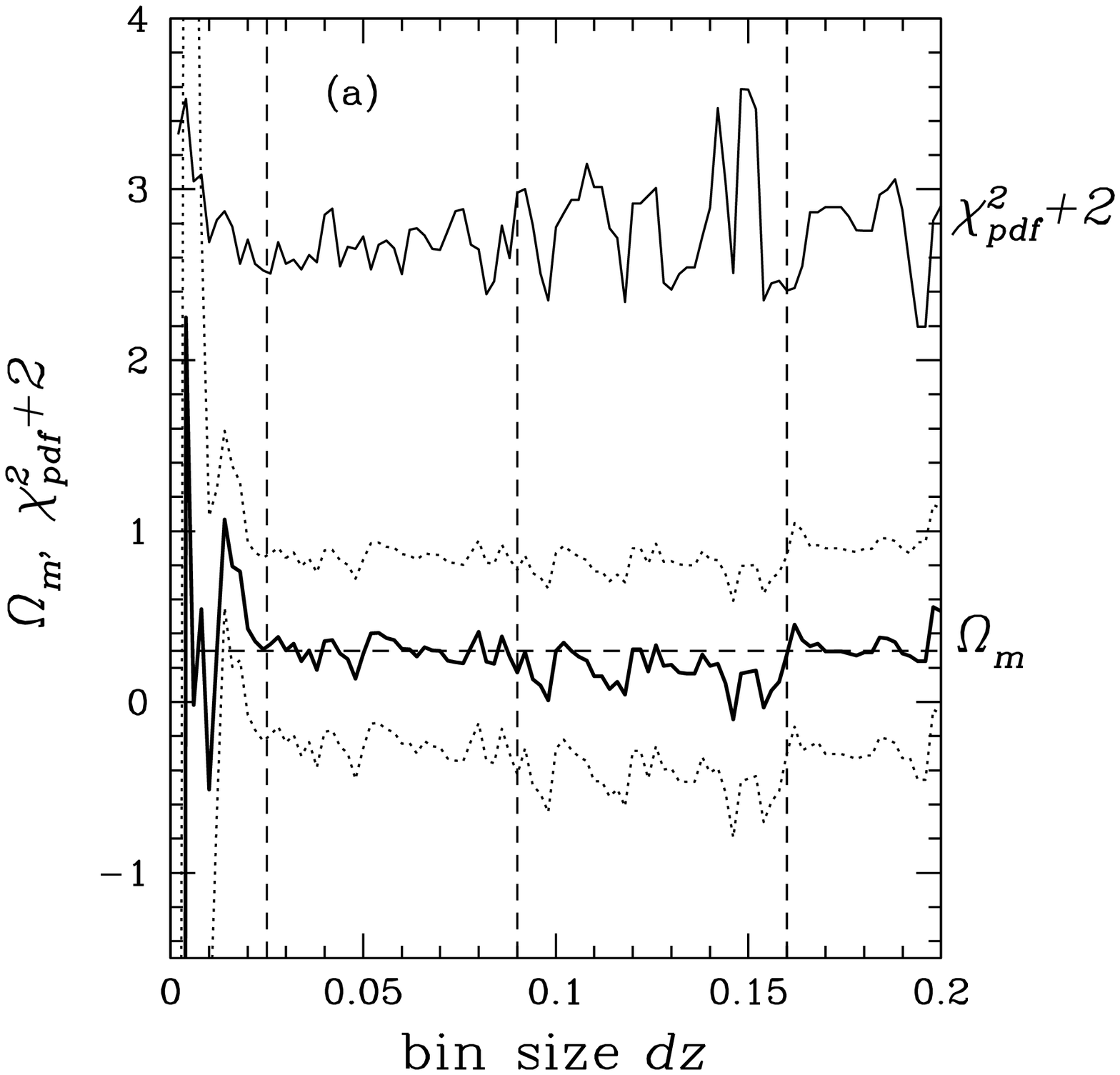]
{The estimated parameters from flux-averaged data
as functions of the size of the redshift bin. 
The thick solid line is the estimated parameter, with 
1$\sigma$ errors marked by
the dotted lines.
The thin solid line is $\chi^2_{\nu}+2$.
(a) $\Omega_m$, (b) $\Omega_{\Lambda}$.}

\clearpage

\setcounter{figure}{0}
\plotone{f1a.eps}
\figcaption[f1a.eps]
{The internal dispersions in the calibrated SN Ia peak absolute magnitudes
from four data sets.
(a) $\mu_0^{MLCS}-\mu_0^p$.}

\setcounter{figure}{0}
\plotone{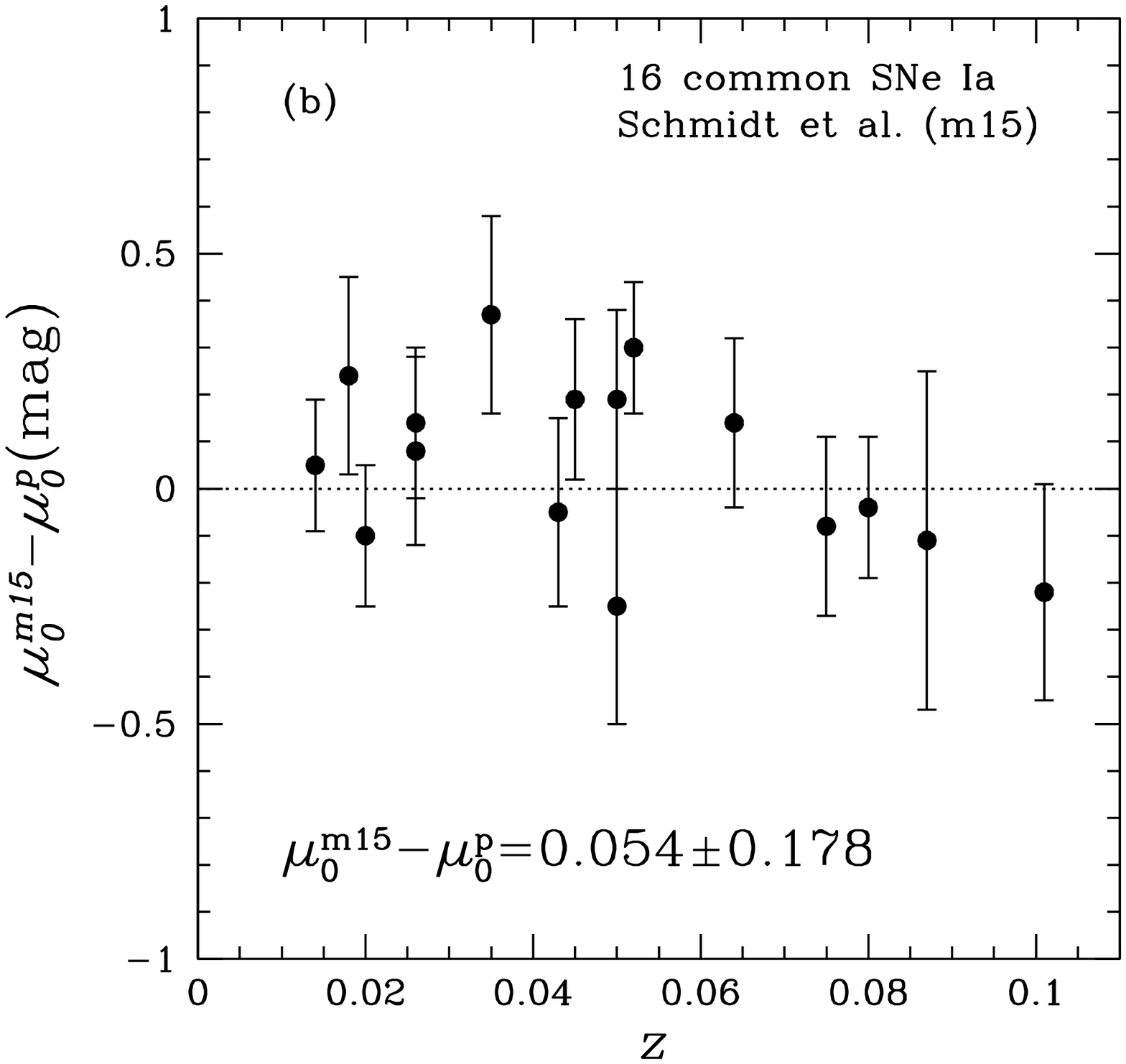}
\figcaption[f1b.eps]
{(b) $\mu_0^{m15}-\mu_0^p$.}

\setcounter{figure}{0}
\plotone{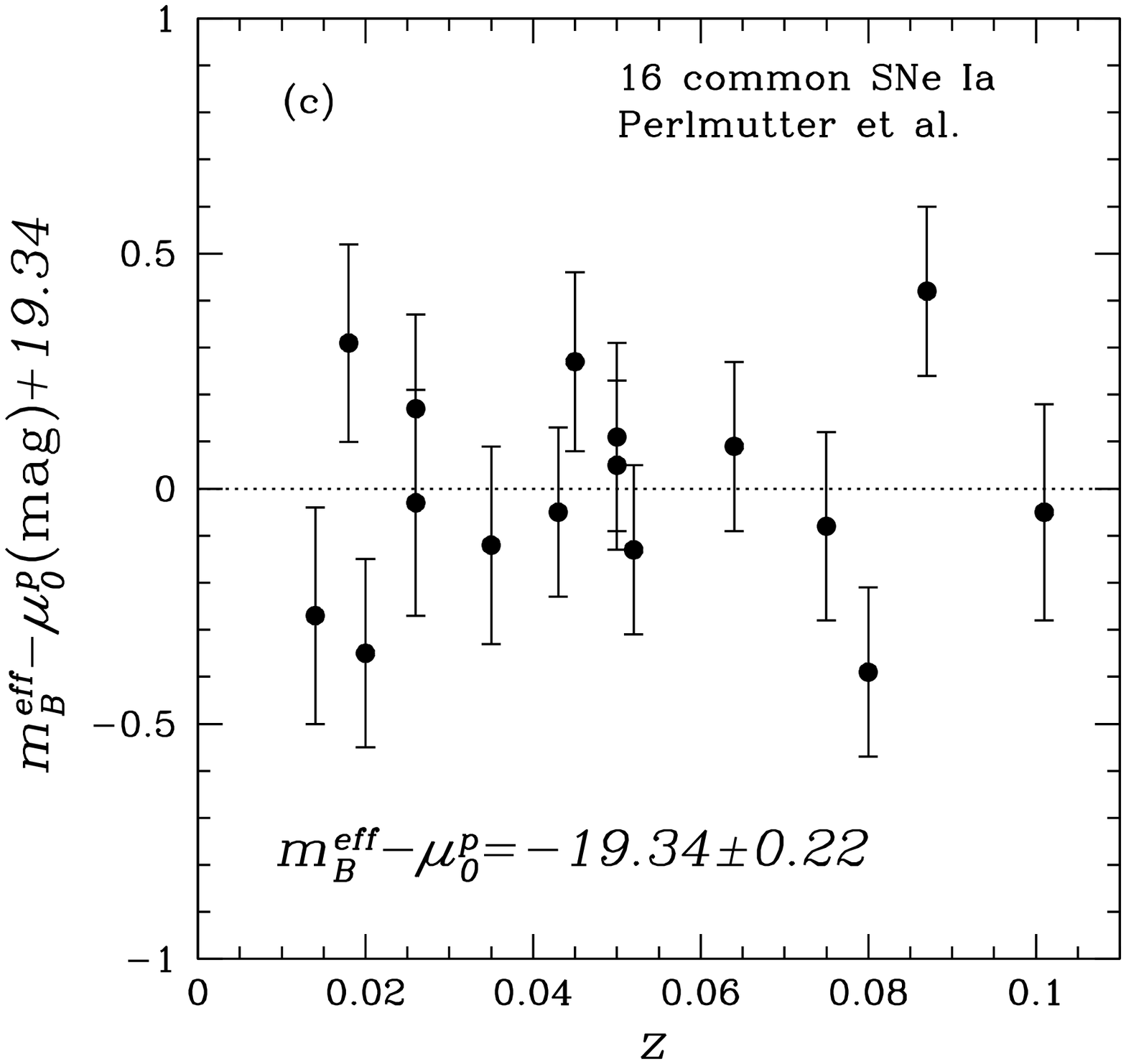}
\figcaption[f1c.eps]
{(c) $m_B^{eff}-\mu_0^p$.}

\setcounter{figure}{0}
\plotone{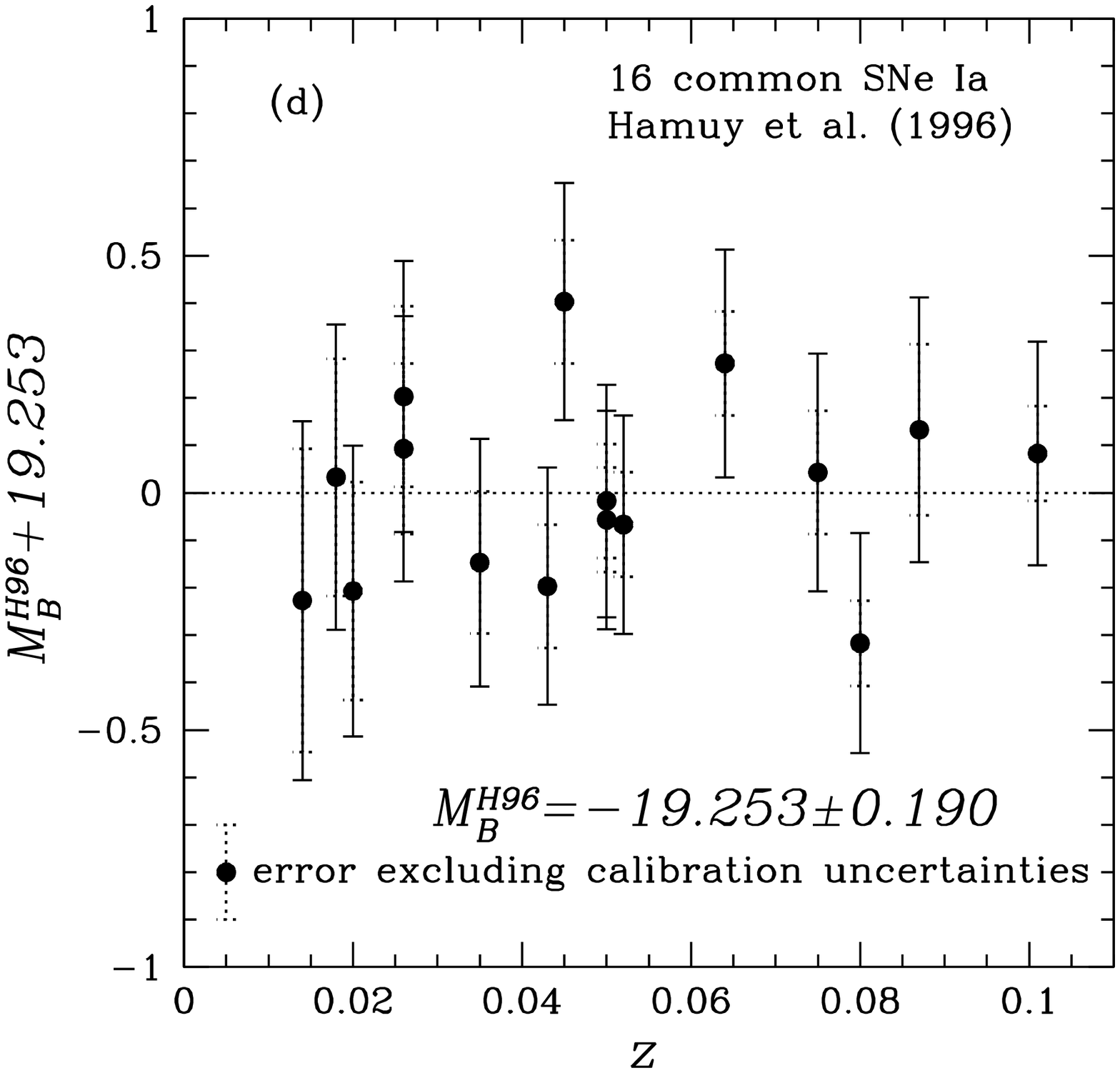}
\figcaption[f1d.eps]
{(d) $M_B^{H96}$.}

\setcounter{figure}{1}
\plotone{f2.eps}
\figcaption[f2.eps]
{The difference between $m_B^{eff}$ (Perlmutter et al. 1999)
and $\mu_0^{MLCS}$ (Riess et al. 1998) for the same 18 SNe Ia.
The error bars are the combined errors in $m_B^{eff}$ and $\mu_0^{MLCS}$.}

\setcounter{figure}{2}
\plotone{f3a.eps}
\figcaption[f3a.eps]
{The magnitude-redshift plots of the combined data set of 
92 SNe Ia. The solid points represent 50 SNe Ia from Schmidt et al. data.
The empty points represent 42 additional SNe Ia from Perlmutter et al.
data. The dashed line is the prediction of OCDM with $\Omega_m=0.2$ 
and $\Omega_{\Lambda}=0$. 
The dotted line is the prediction of SCDM with $\Omega_m=1$
and $\Omega_{\Lambda}=0$. The two thick solid lines are predictions
of $\Lambda$CDM, with ($\Omega_m, \,\Omega_{\Lambda})=(0.3,\,0.7)$,
($\Omega_m, \,\Omega_{\Lambda})=(0.2,\,0.8)$ respectively.
The thin solid line is SCDM with $(1+z)$ dimming of SN Ia peak luminosity.
(a) The distance modulus versus redshift.}

\setcounter{figure}{2}
\plotone{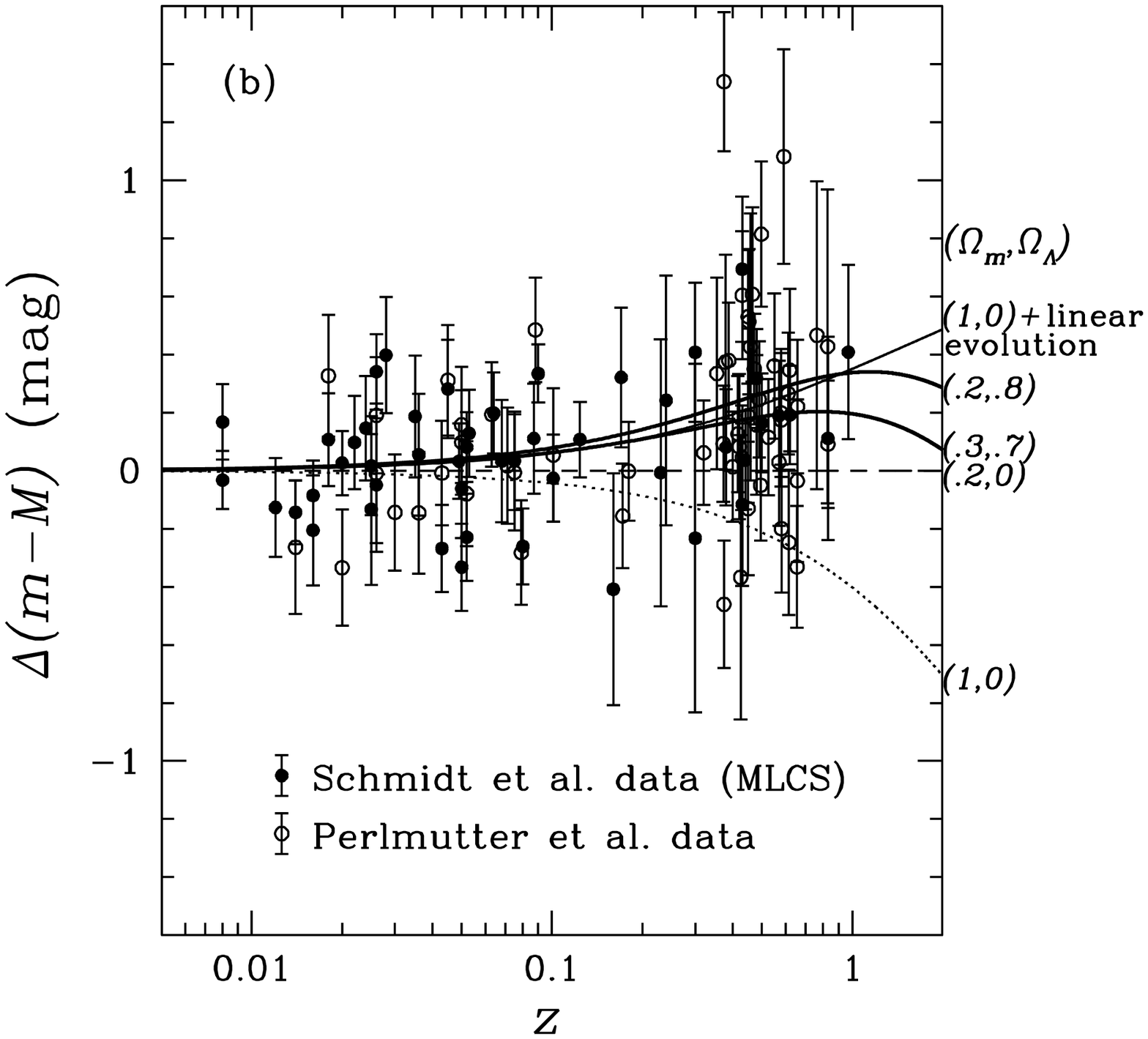}
\figcaption[f3b.eps]
{(b) The distance modulus relative to the prediction
of the OCDM model.}

\plotone{f4.eps}
\figcaption[f4.eps]
{The 68.3\% and 95.4\% confidence contours in the 
$\Omega_{\Lambda}-\Omega_m$ plane. The dotted lines represent
the Schmidt et al. data (50 SNe Ia), the dashed lines represent
the Perlmutter et al. data (60 SNe Ia), and the solid lines represent
the combined data (92 SNe Ia).}

\plotone{f5a.eps}
\figcaption[f5a.eps]
{The magnitude-redshift plots of the binned data for 92 SNe Ia,
with redshift bin $\Delta z=0.05$. The line types are the same as in Fig.3.
(a) The distance modulus versus redshift.}

\setcounter{figure}{4}
\plotone{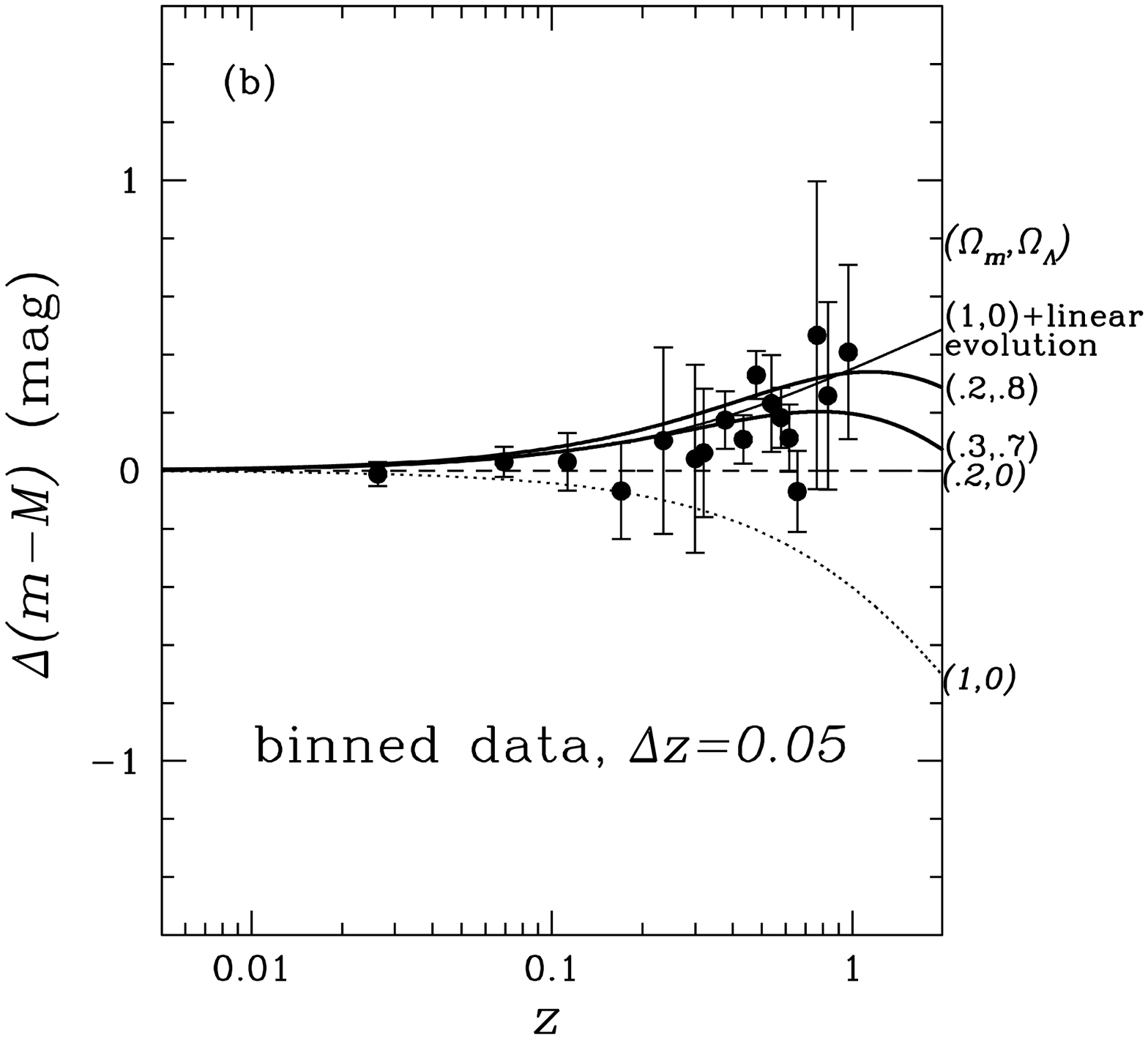}
\figcaption[f5b.eps]
{(b) The distance modulus relative to the prediction
of the open cold dark matter model (OCDM) with $\Omega_m=0.2$
and $\Omega_{\Lambda}=0$. }

\plotone{f6a.eps}
\figcaption[f6a.eps]
{The same as Fig.5, but with redshift bin $\Delta z=0.1$.
(a) The distance modulus versus redshift.}

\setcounter{figure}{5}
\plotone{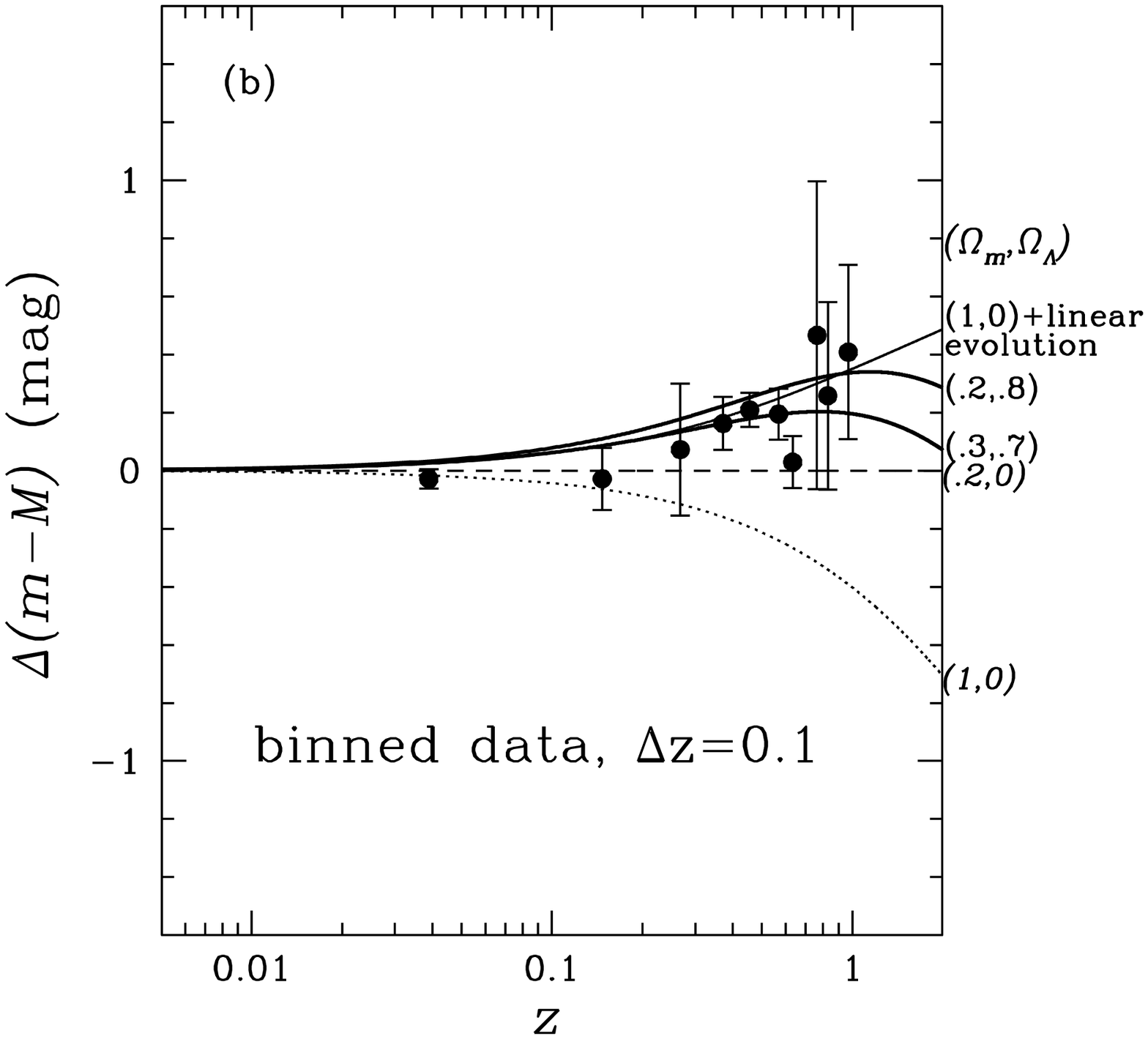}
\figcaption[f6b.eps]
{(b) The distance modulus relative to the prediction
of the open cold dark matter model (OCDM) with $\Omega_m=0.2$
and $\Omega_{\Lambda}=0$. }

\plotone{f7.eps}
\figcaption[f7.eps]
{The 68.3\% and 95.4\% confidence contours in the 
$\Omega_{\Lambda}-\Omega_m$ plane. The solid lines represent the unbinned
data, the dotted lines represent the binned data with 
redshift bin $\Delta z=0.05$; the dashed lines represent
the binned data with redshift bin $\Delta z=0.1$.}

\plotone{f8a.eps}
\figcaption[f8a.eps]
{The estimated parameters from flux-averaged data
as functions of the size of the redshift bin. 
The thick solid line is the estimated parameter, with 
1$\sigma$ errors marked by
the dotted lines.
The thin solid line is $\chi^2_{\nu}+2$.
(a) $\Omega_m$.}

\setcounter{figure}{7}

\plotone{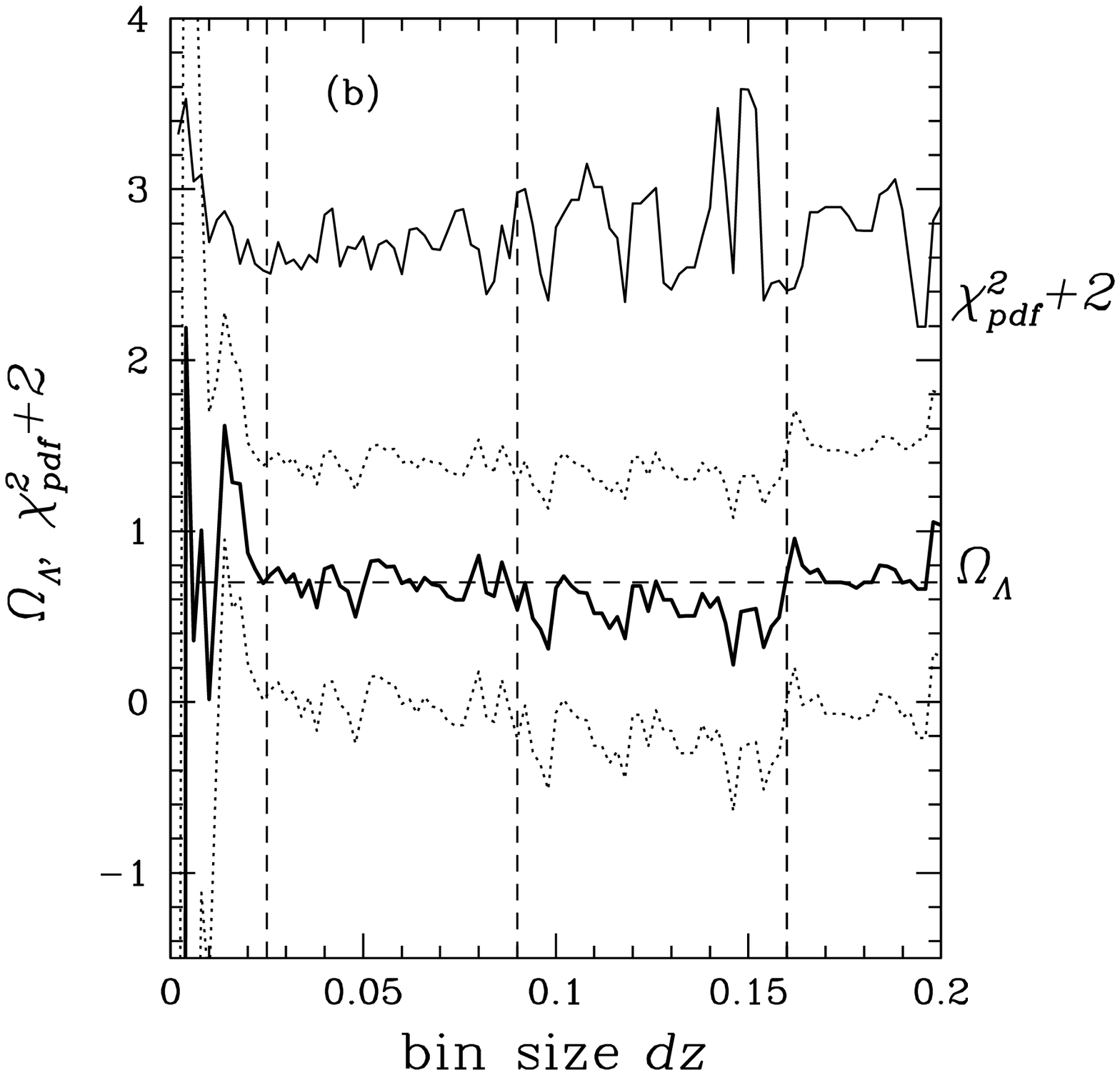}
\figcaption[f8b.eps]
{(b) $\Omega_{\Lambda}$.}

\end{document}